\documentclass[twocolumn,showpacs,preprintnumbers,amsmath,amssymb]{revtex4-1}
\usepackage{amsmath}
 \usepackage{lipsum}
\usepackage{epsfig}
\usepackage{graphicx}
\usepackage{dcolumn}
\usepackage{bm}
\usepackage{wrapfig}
\usepackage{xcolor}
\usepackage{hyperref}
\definecolor{linkcolor}{HTML}{799B03}
\begin{document}
\title{Chimeras and solitary states in 3D oscillator networks  with inertia}
\author{V. Maistrenko$^{1}$, O. Sudakov$^{1,2}$, O. Osiv$^{1}$}
\address{$^{1}$Scientific Center for Medical and Biotechnical Research, \\
NAS of Ukraine, 54, Volodymyrs'ka Str., Kyiv 01030, Ukraine  \\
$^{2}$Taras Shevchenko National University of Kyiv, 60, Volodymyrs'ka Str., Kyiv 01030, Ukraine}

\begin{abstract}
We report the diversity of scroll wave chimeras in the three-dimensional (3D) Kuramoto model with inertia for $N^{3}$ identical phase oscillators placed in a unit 3D cube with periodic boundary conditions. In the considered model with inertia, we have found  patterns which do not exist in this system without inertia. In particular, a scroll ring chimera is obtained from random initial conditions. In contrast to this system without inertia, where all chimera states have incoherent inner parts, these states can have partially coherent or fully coherent inner parts as exemplified by a scroll  ring  chimera. Solitary states exist in the considered model as separate states or can coexist with scroll wave chimeras in the oscillatory space. 
We also propose a method of construction of 3D images using solitary states as solutions of the  3D Kuramoto model with inertia.

\end{abstract}
\maketitle

{\bf Chimera states as a phenomenon of coexistence of coherence and incoherence patterns are well known in nonlinear science.  This phenomenon appears in models that describe the basic properties of a collective dynamics in various real physical systems.
In our study within 3D oscillatory networks with inertia, we investigate the appearance of chimeras and analyze their properties by the example of a scroll  ring  chimera obtained from random initial conditions.
Our results show that, due to the introduction of the inertia, the 3D chimeras are stable despite perturbations of the initial conditions and can be surrounded by solitary oscillators.}

Chimera states~\cite{k2002,kb2002,as2004}  
as a dynamical phenomenon in arrays of non-locally coupled
oscillators, that displays a spatio-temporal pattern of co-existing coherence and incoherence,
 have been elaborately investigated during the past decade in a wide range of systems. 

A number of illustrious articles are devoted to the theoretical and experimental studies of chimeras. Most of them deal with {\it one-dimensional} models.
Nonetheless,  this novel approach was extended to the {\it two-dimensional} networks of oscillators.   
Among other factors, a new class of chimera states called the spiral wave chimeras has been introduced in~\cite{ks2003}.  This kind of spatio-temporal behavior is characterized by the standard 2D spiraling and, moreover, by a finite-size incoherent core (see, e.g.,~\cite{k2004,mls2010,OWYYS2012,pa2013,XKK2015,L2017,OWK2018,TRT2018,OK2019,T2019}).

As for the 3D case, we mention a few papers about  chimera states in the model of coupled phase oscillators in a { \it three-dimensional} grid topology\cite{msom2015,ld2016,msom2017,khp2018,ok2019,kbgl2019}.
The first evidence of chimera states in 3D was reported in 2015 in~\cite{msom2015} for the Kuramoto model of coupled phase oscillators in a 3D grid topology with piecewise constant oscillator's coupling. Just this kind of coupling produces a rich variety of chimera states in the 1D Kuramoto model  with respect to cosine and exponential couplings ~\cite{mvslm2014}.
In ~\cite{msom2015} the 
3D oscillating chimera states, i.e., those with no spiraling of the coherent region (incoherent and coherent balls, tubes, crosses, layers), and  spiral-shaped rotating chimeras called {\it scroll wave chimeras} including  incoherent rolls of different modalities in the coherent spiral arms were obtained.
Some time later, two new kinds of scroll wave chimeras, Hopf link and trefoil,  with linked and knotted incoherent regions were detected in~\cite{ld2016}. 
The theoretical confirmation of the existence of a few kinds of  3D chimera states in the Kuramoto model  with the cosine coupling of oscillators was done in 2019 in~\cite{ok2019}.

In the present paper, we study the appearance of 3D chimera states in the Kuramoto model of coupled phase oscillators in the 3D grid topology with inertia:

\vspace*{-0.6cm}
\begin{widetext}
\begin{eqnarray}
m\ddot{\varphi}_{ijk} + \epsilon \dot{\varphi}_{ijk}  = 
 \frac{\mu}{|B_{P}(i,j,k)|} \sum\limits_{(i^{\prime},j^{\prime},k^{\prime})\in B_{P}(i,j,k) }\sin(\varphi_{i^{\prime}j^{\prime}k^{\prime}} - \varphi_{ijk}- \alpha),
\end{eqnarray}
\vspace*{-0.3cm}
\end{widetext}
\vspace*{-0.6cm}
where   $i,j,k = 1, ... , N$, $\varphi_{ijk}$ are phase variables, and indices $i,j,k$ are periodic modulo $N$.   The coupling  is  assumed long-ranged and isotropic:  each oscillator $\varphi_{ijk}$ is coupled with equal strength $\mu$ to all its nearest neighbors $\varphi_{i^{\prime}j^{\prime}k^{\prime}}$  within  a ball of radius $P$, i.e., to those falling in the neighborhood 
$$  \small B_{P}(i,j,k):=\{ (i^{\prime},j^{\prime},k^{\prime}){:} (i^{\prime}-i)^{2}+(j^{\prime}-j)^{2}+(k^{\prime}-k)^{2}\le P^{2}\},$$
where the distances 
are calculated taking into account the periodic boundary conditions of the network.
The phase lag parameter $\alpha$ is selected from  the range $[0, \pi/2]$.  
The  relative coupling radius $r=P/N,$ varies from $1/N$ (local coupling) to $0.5$ (close to the global coupling). 

\begin{figure*}[ht!]
\vspace*{-0.4cm}
 \center{\includegraphics[width=0.225\linewidth]{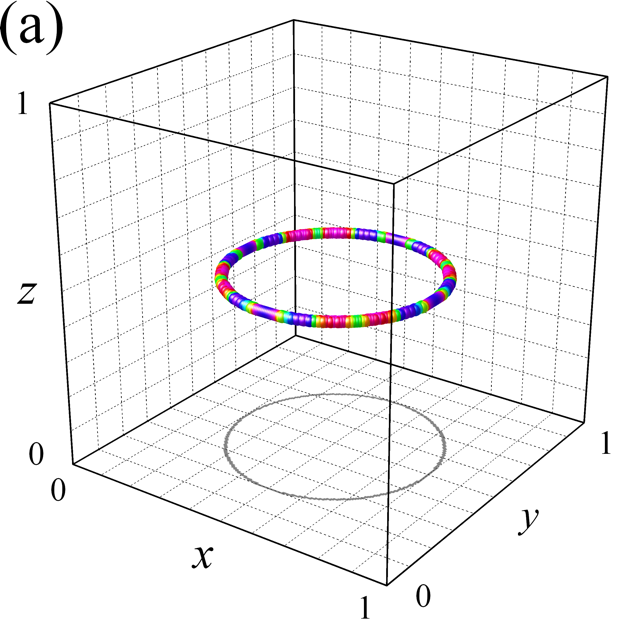}  \  \   
  \includegraphics[width=0.225\linewidth]{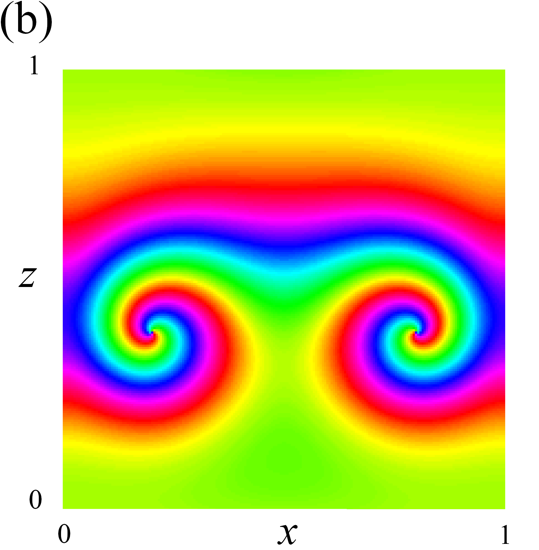}  \  \  
  \includegraphics[width=0.225\linewidth]{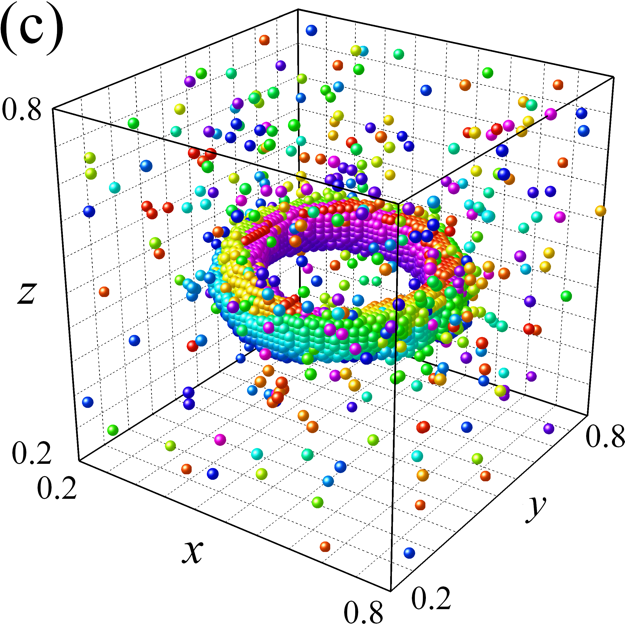} \    \  
  \includegraphics[width=0.225\linewidth]{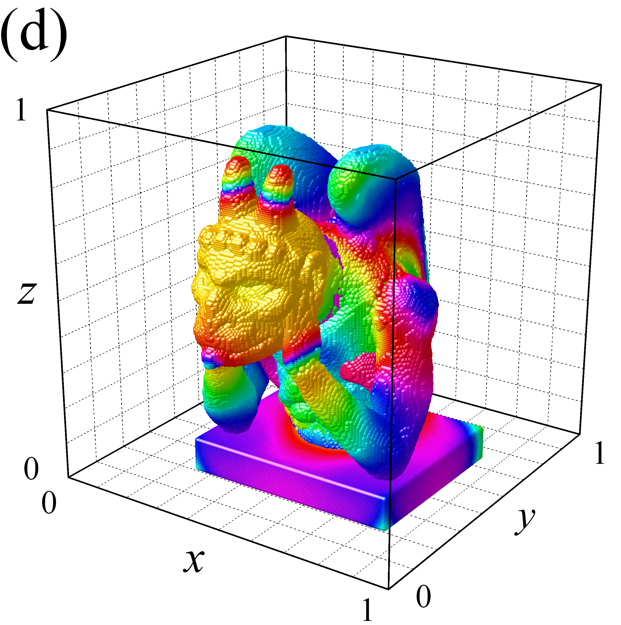} \          \ 
  \includegraphics[width=0.03\linewidth]{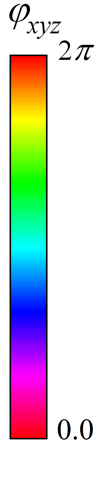}} 
\vspace*{-0.4cm}
 \caption{  Phase snapshots of patterns  in model (1): (a) - scroll ring, (b) - cross-sections of (a) at $y=0.5 $ ($\alpha=0.38, \mu=0.02, r=0.01, N=200$), (c) -  scroll ring  with solitary cloud ($\alpha=0.4, \mu=0.1, r=0.04, N=100$), (d) - 3D image of a chimera  sculpture generated by solitary states ($\alpha=0.35, \mu=0.1, r=0.15, N=200$); $ \epsilon=0.05$. 
 Coordinates  $x=i/N,  y=j/N,  z=k/N$. } 
\label{fig:1}
\end{figure*}

The parameter $\mu$ is the oscillator coupling strength, and $\epsilon$ is the damping  coefficient.
 The parameter $m$ is the mass. In the case $m=0,$ Eq. (1) is transformed into the pure 3D Kuramoto model  without inertia.
We put $m=1$ without any loss of generality.

We study  the appearance of a variety of scroll wave chimeras which do not exist in the model without inertia. Their properties are analyzed  through the example of scroll ring chimera states (Fig. 1(a,b)).  The scroll wave chimera obtained from random initial conditions looks like a mis-shapen scroll ring (Fig. 2) and will be referred hereinafter as a {\it scroll ring chimera state}  \cite{w1973,bb2014,tes2015,ld2016}. 

Scroll wave chimera states in system (1) without inertia  for piecewise constant oscillator's coupling always have incoherent inner parts \cite{msom2015,ld2016,msom2017}.
In contrast, the scroll wave chimera in system (1) with inertia can have partially coherent or fully coherent inner parts.
The diversity of scroll wave chimeras and the impact of inertia on scroll wave chimeras in a system without inertia are  analyzed.

We demonstrate that the 3D scroll wave chimeras can be surrounded by solitary oscillators ~\cite{JMK2015,JBLDKM2018}  (like a ``solitary cloud'') (Fig. 1 (c)). 
 In~\cite{msm2019}, the spiral wave chimeras with solitary clouds have been studied in the case of the 2D Kuramoto model with inertia.

Finally  we propose a method of construction of 3D images using solitary states as solutions of the  inertial system (1). An example of a chimera sculpture from the Notre-Dame  Cathedral  is shown in Fig. 1(d).

Numerical simulations were performed on the base of the Runge--Kutta solver DOPRI5 on a computer cluster CHIMERA, http://nll.biomed.kiev.ua/cluster,  with graphics processing units~\cite{sls2011,scm2017}.
Trajectories were computed and analyzed for system (1) with $N = 50, 100, 200$ (0.125, 1, 8 mln oscillators, respectively). Coupling radius $P$  was selected from  the range $[1, 50]$. The random initial conditions were chosen from  the phase range $[0,2\pi]$ and the frequency range $[-\mu/\epsilon, \mu/\epsilon]$, independently for each oscillator.

During the simulation  of Eq. (1) without inertia  ($m=0$), we typically observed a birth of scroll ring chimeras from random initial conditions. But in all such cases, the scroll ring  appears to be unstable and is destroyed rather rapidly after the start of a simulation.  A similar phenomenon was reported in \cite{ld2016}.

In the case of the model with inertia ($m=1$), the random initial condition generates a stable scroll ring  chimera which survives, as far as we simulate in time till $t=10^{5}$ with the parameter values $\alpha=0.2, r=0.03$, $ \mu=0.1, \epsilon=0.05, N=100$.

Figure 2 \href{video 1.avi} {(Multimedia view)}
illustrates the time evolution of the random trajectory. At the time $t \approx 2500,$ two scroll rings have been generated by the random chaotic behavior of oscillators. They exist during the time interval $t \approx (2500-3900)$ (Fig. 2 (a)). Then one ring shrinks and disappears, leading to homogeneous
oscillations, but the second  ring still survives (Fig. 2 (b)), as far as we simulated (till $10^{5}$ time units), and is fixed in the oscillatory space. Its cross-section  along $y=0.5 $ is presented in Fig. 2(c).
 
\begin{figure}[ht!]
\vspace*{-0.3cm}
 \center{\includegraphics[width=0.306\linewidth]{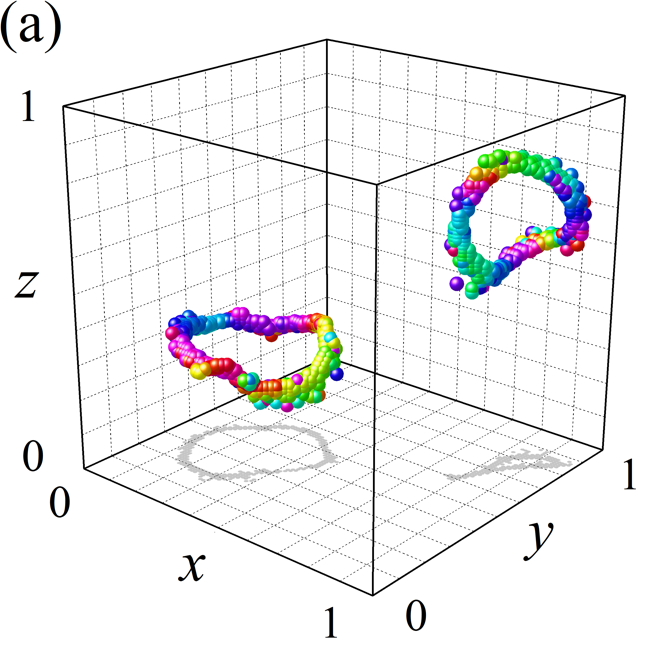}    
  \includegraphics[width=0.306\linewidth]{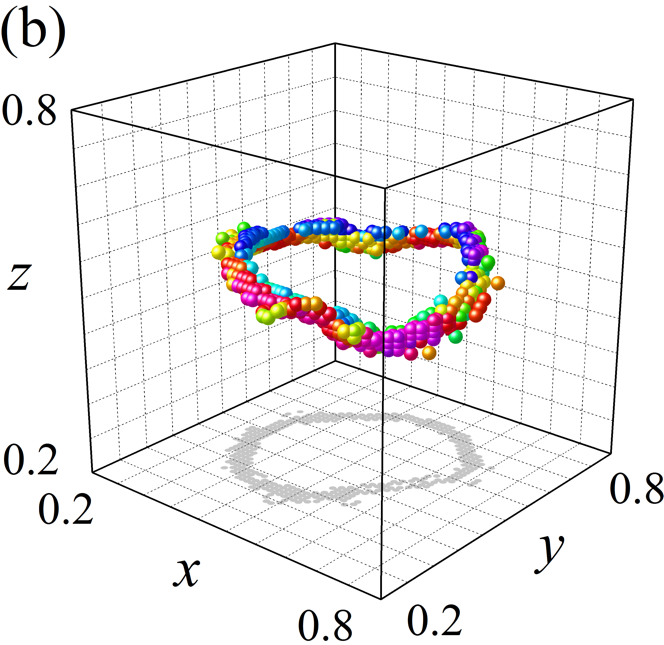} 
  \includegraphics[width=0.29\linewidth]{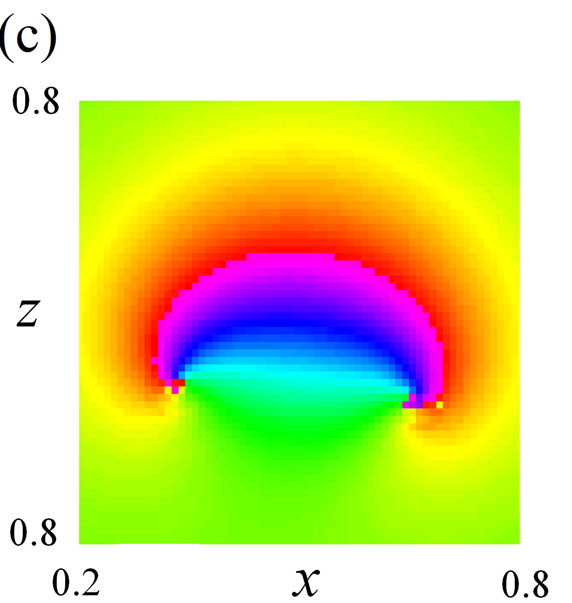}  
  \includegraphics[width=0.062\linewidth]{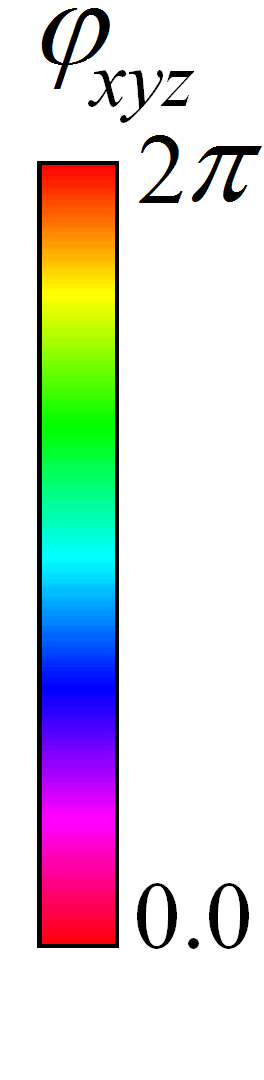}} 
\vspace*{-0.6cm}
 \caption{  Generated  scroll ring chimera state from random initial conditions. Phase snapshots at $t=2500$ (a), $t=4000$ (b), cross-section of (b) along $y=0.5 $ (c). 
Parameters $\alpha=0.2, r=0.03$, $ \mu=0.1, \epsilon=0.05, N=100$. 
 Coordinates  $x=i/N,  y=j/N,  z=k/N$. \href{video 1.avi} {(Multimedia view)}.} 
\label{fig:2}
\end{figure}

\vspace*{-0.4cm}
Other patterns in model (1) were detected from random initial condition as well. 
The pattern is considered stable, if it remained stationary for more than 1000 time units.
After identifying these patterns, the parameter region for their existence was explored by the standard continuation method with changing parameter values. 

 \begin{figure*}[ht!]
\vspace*{-0.8cm}
\includegraphics[width=0.35\linewidth]{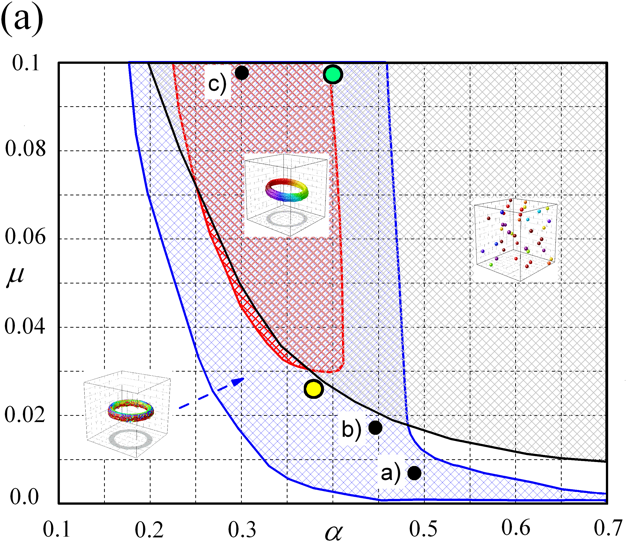} 
\hspace*{2.0cm} 
\includegraphics[width=0.35\linewidth]{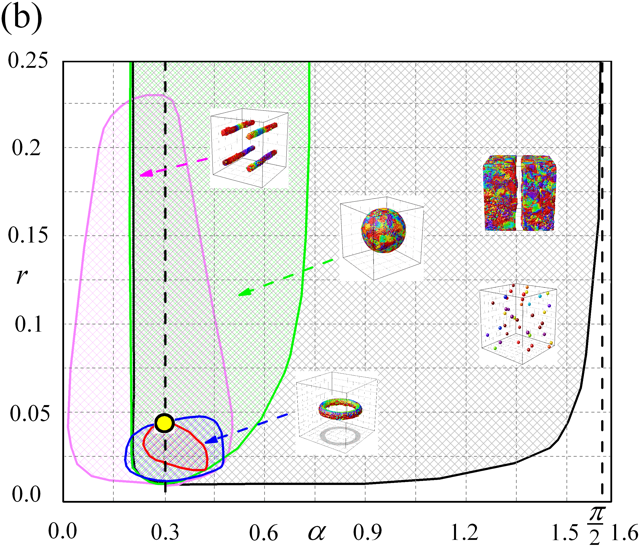} 
\vspace*{-0.4cm}
\caption {Parameter regions  for chimeras and solitary states in the parameter planes: 
(a) - $(\alpha, \mu)$  $( r=0.04)$, (b) - $(\alpha, r)$ $(\mu=0.1)$.
Blue - scroll ring with incoherent or partially coherent inner parts, red - coherent ring,  green - sphere,   magenta - 4 scroll wave rods, gray - solitary states. Snapshots of typical states are shown in the insets. $\epsilon=0.05, N=100$.}
  \label{fig:3}
\end{figure*}

\begin{figure*}[th!]
\vspace*{-0.3cm}
\includegraphics[width=0.22\linewidth]{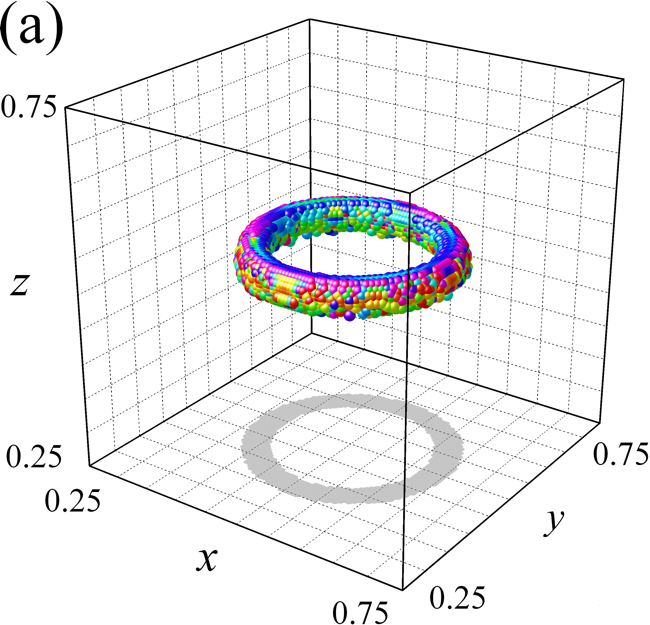}   \hspace*{0.19cm}
   \includegraphics[width=0.03\linewidth]{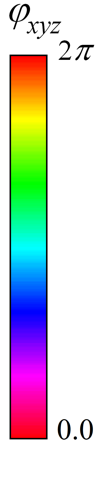}
\includegraphics[width=0.215\linewidth]{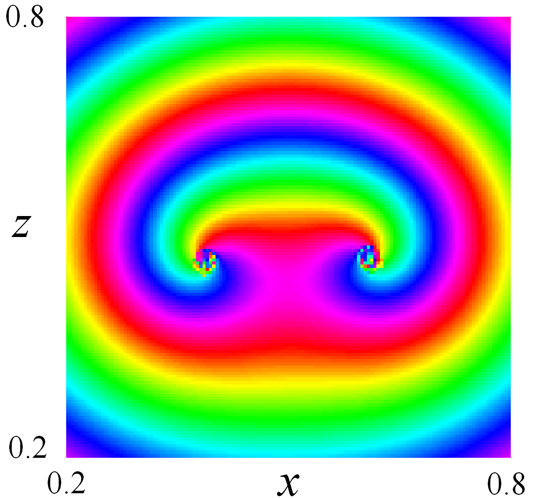} \hspace*{0.05cm}
\includegraphics[width=0.21\linewidth]{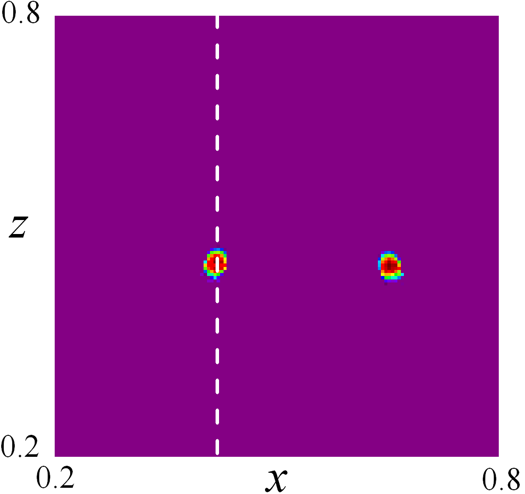} 
  \includegraphics[width=0.039\linewidth]{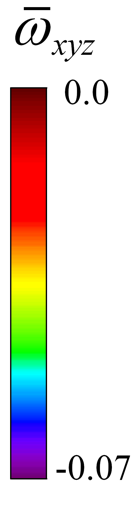}
\includegraphics[width=0.23\linewidth]{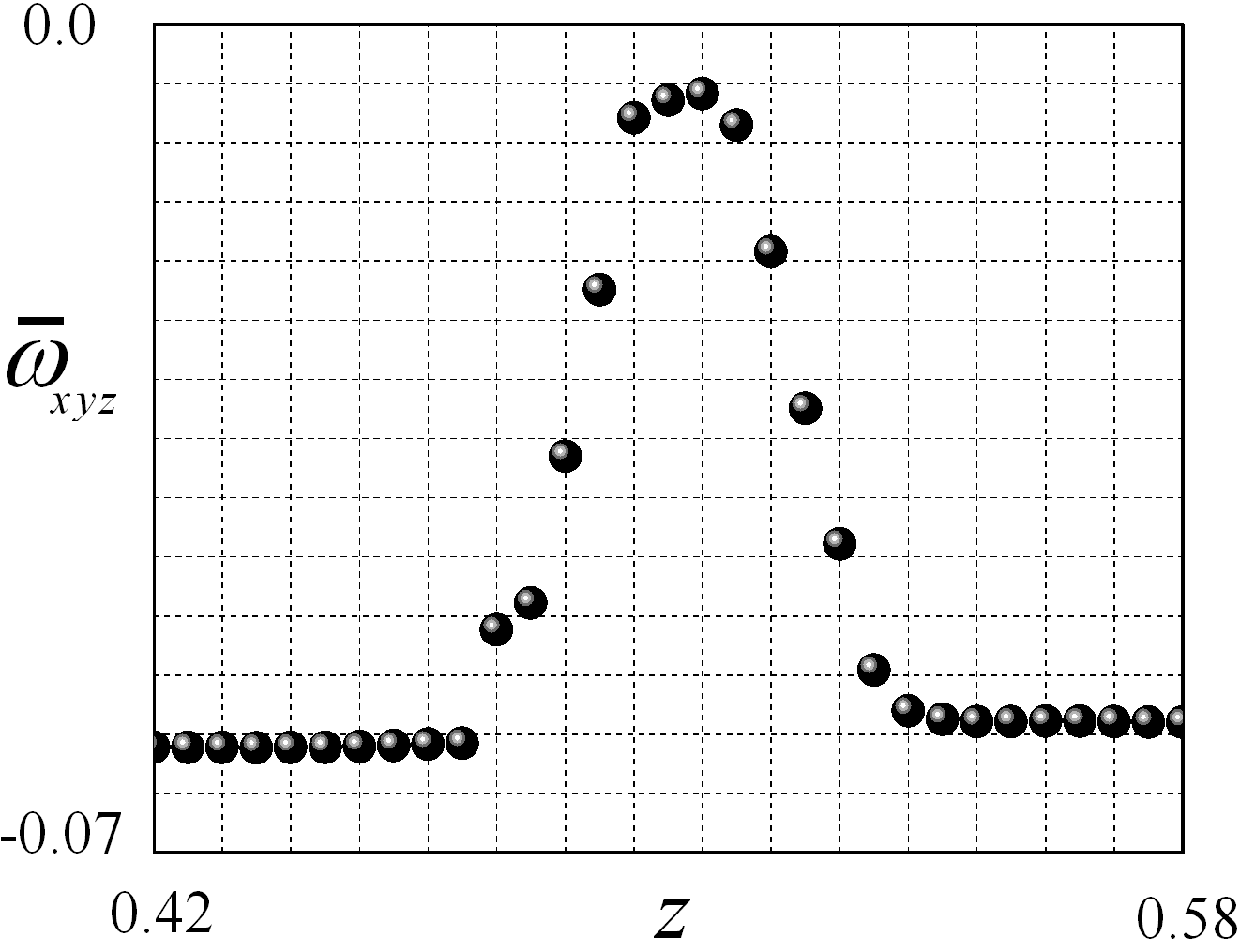}  \\
\vspace*{0.1cm}
  \includegraphics[width=0.22\linewidth]{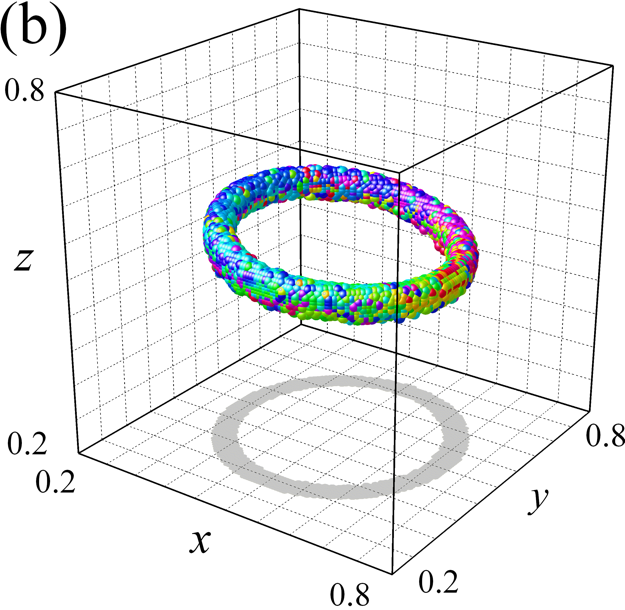}  \hspace*{0.19cm}
   \includegraphics[width=0.03\linewidth]{Faza-insert-F4.png}
   \includegraphics[width=0.215\linewidth]{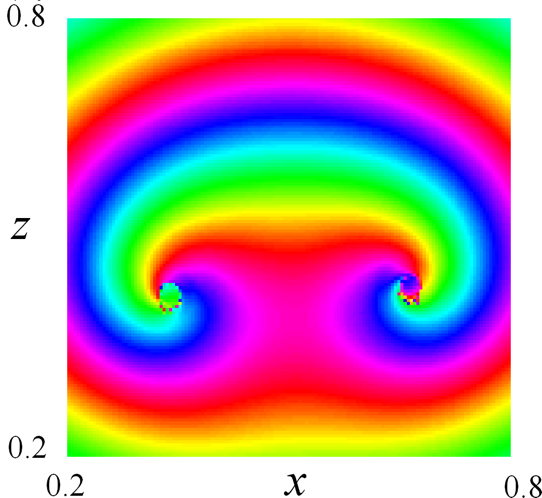} \hspace*{0.05cm}
  \includegraphics[width=0.21\linewidth]{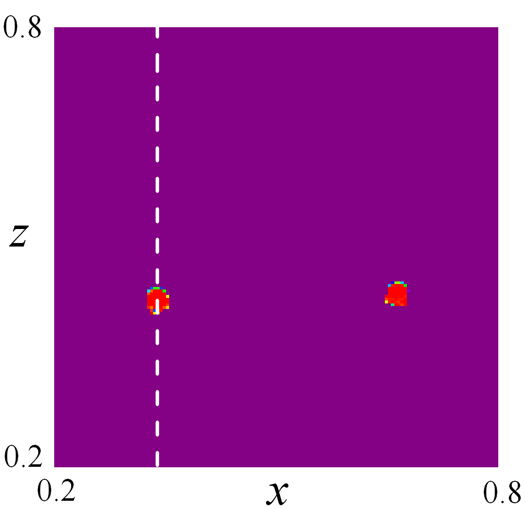}  
  \includegraphics[width=0.039\linewidth]{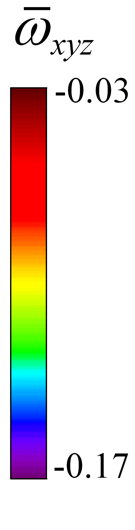}
  \includegraphics[width=0.23\linewidth]{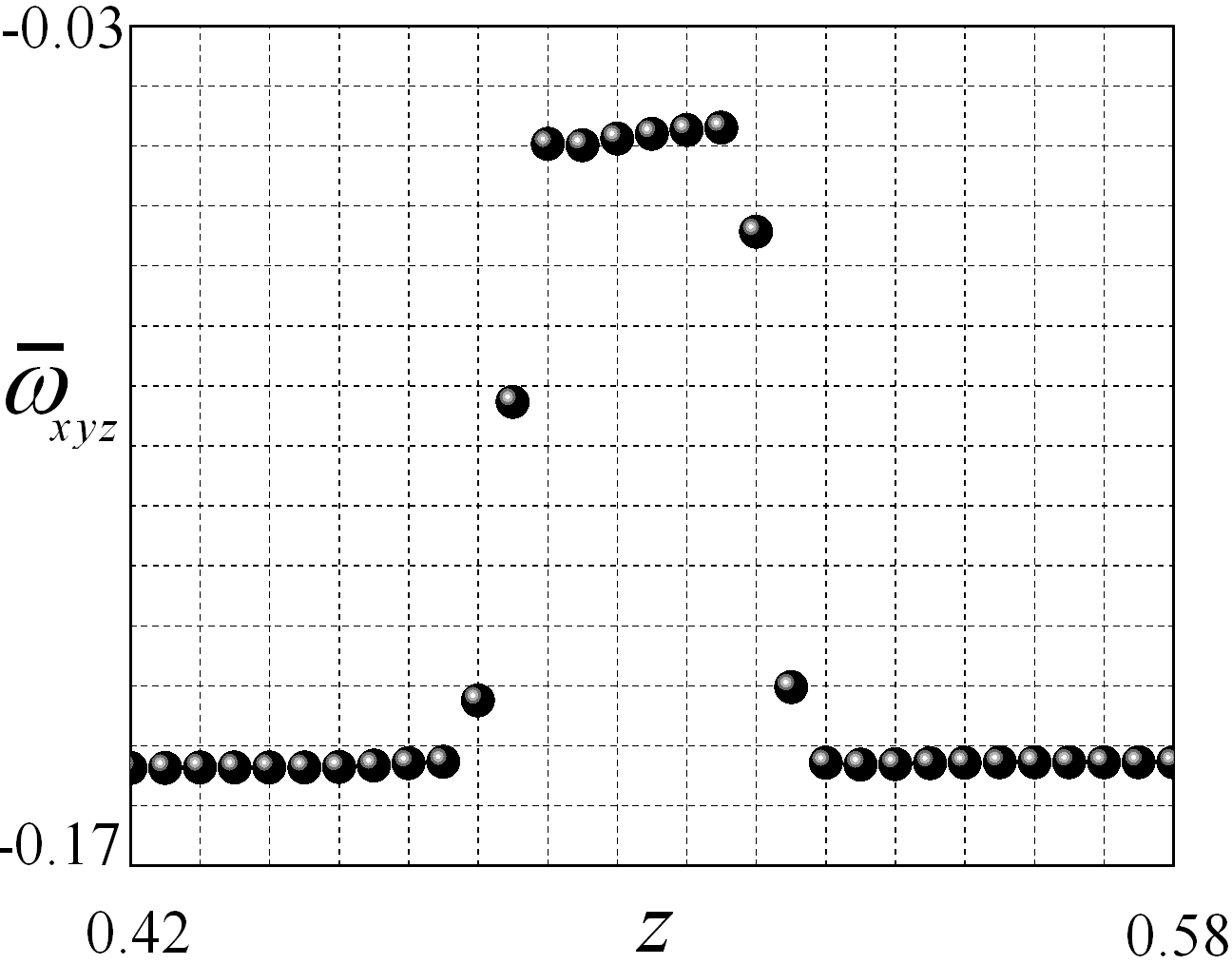}  \\ 
\vspace*{0.2cm}
\includegraphics[width=0.222\linewidth]{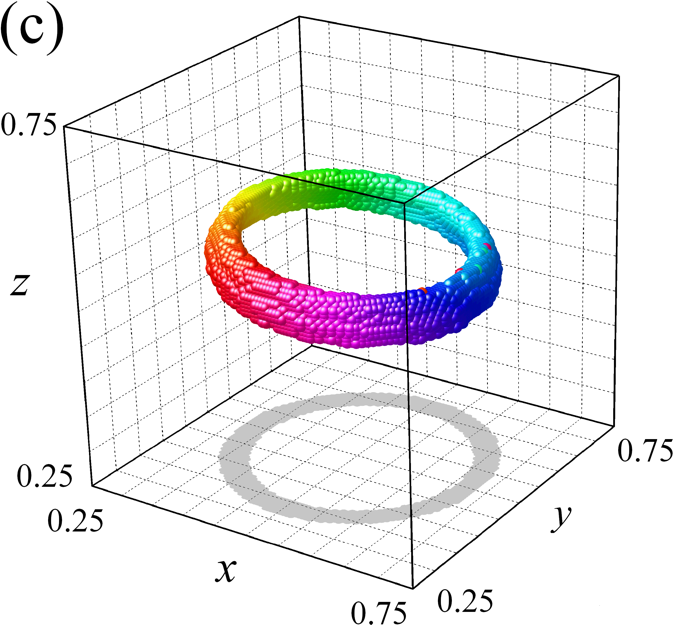}  \hspace*{0.19cm}
 \includegraphics[width=0.03\linewidth]{Faza-insert-F4.png}
  \includegraphics[width=0.215\linewidth]{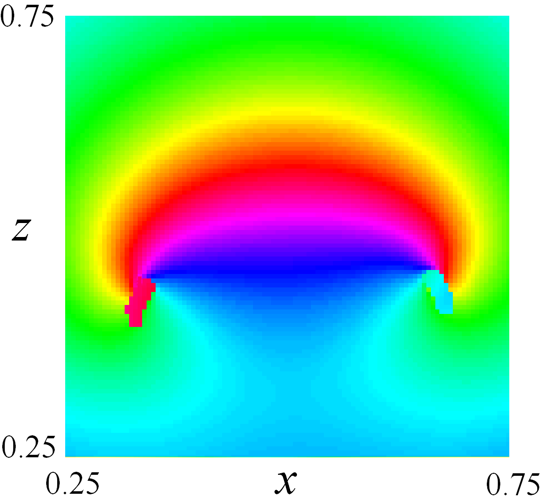} \hspace*{0.05cm}
  \includegraphics[width=0.21\linewidth]{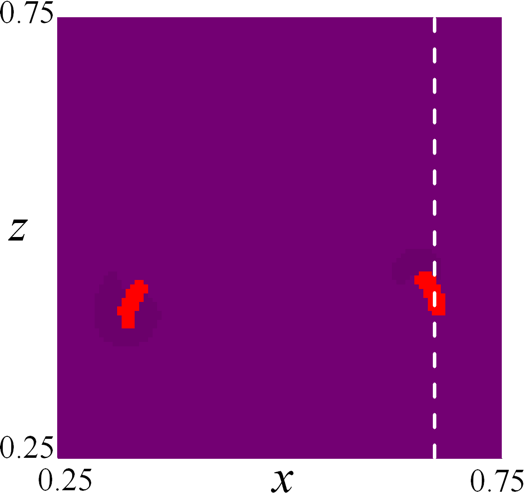} 
  \includegraphics[width=0.039\linewidth]{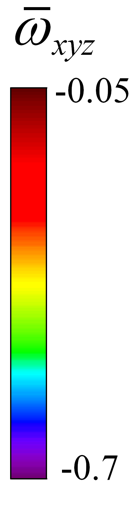}
  \includegraphics[width=0.23\linewidth]{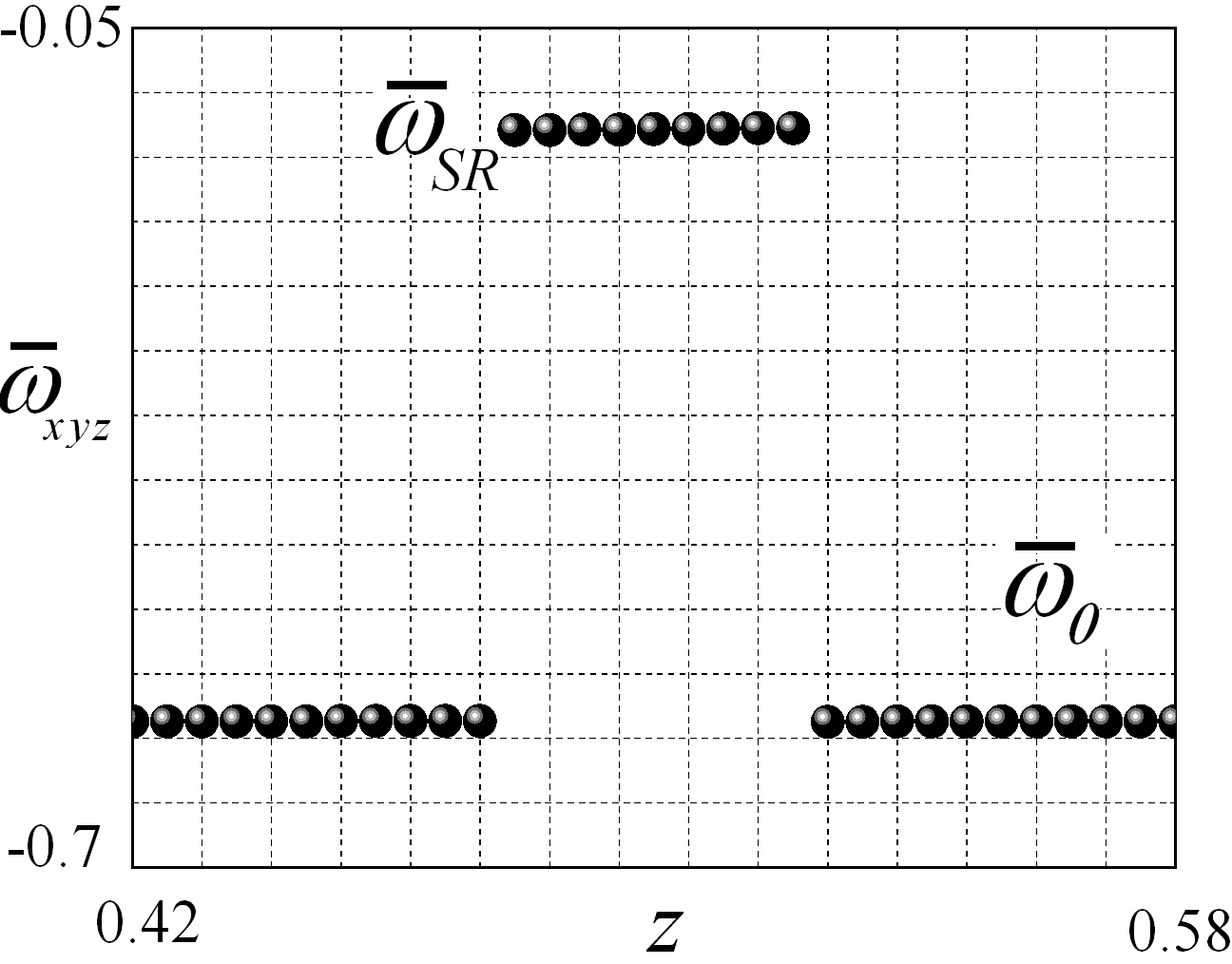} 
\vspace*{-0.6cm}
 \caption{Examples of scroll ring chimeras. 
Left column -  phase snapshots, next two  columns - cross-sections of the phase and
average frequencies at the centers of the scroll rings, right column - double cross-sections of average frequencies  along the white dashed line on the previous column.
  (a) - incoherent scroll ring chimera ($\alpha=0.475, \mu=0.007$) \href{video 2.avi} {(Multimedia view)}, (b) - scroll ring chimera with partially coherent inner part ($\alpha=0.436, \mu=0.019$);  (c) - coherent scroll ring. Parameters  
   $r=0.04, \epsilon=0.05, N=200$. Simulation time $t=10^4$. Frequency averaging interval  $\Delta T = 1000$.}
\label{fig:4}
\end{figure*}

In Fig. 3, the parameter regions  for a scroll ring chimera and for a few other patterns in  $(\alpha, \mu)$ (a) and $(\alpha, r)$ (b)  parameter planes  are presented. The blue region  in Fig. 3(a) corresponds to the scroll ring chimera with incoherent or partially coherent inner part, red - scroll ring with completely coherent inner part. The stability region for solitary states, which are considered in detail below, is hatched by gray. 
Typical shapes of the chimera states are shown in insets.
As our simulation shows, the scroll ring chimera states exist for any infinitely small coupling strength $\mu>0$. Crossing the left and left bottom sides of the scroll ring stability region, all oscillators are synchronized.
Figure 3(b)  also illustrates the parameter regions, but in $(\alpha, r)$ parameter plane with additional regions for a sphere or ball (green) and  4 scroll wave rods (magenta). 
Here, the  lower bound of the scroll ring region starts from the coupling radius $P=1$ (local coupling).
Crossing the left and bottom sides of this region, all oscillators are synchronized, and the ring is vanished. After crossing the right side of the region, the rings are destroyed with the generation of the chaotic oscillatory behavior. 

\begin{figure*}[ht!]
\vspace*{-0.5cm}
 \center{
  \includegraphics[width=0.22\linewidth]{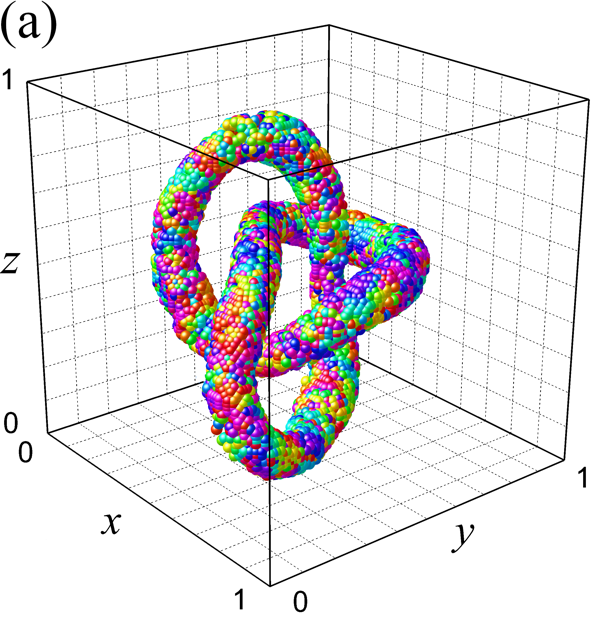}  \ \   
  \includegraphics[width=0.22\linewidth]{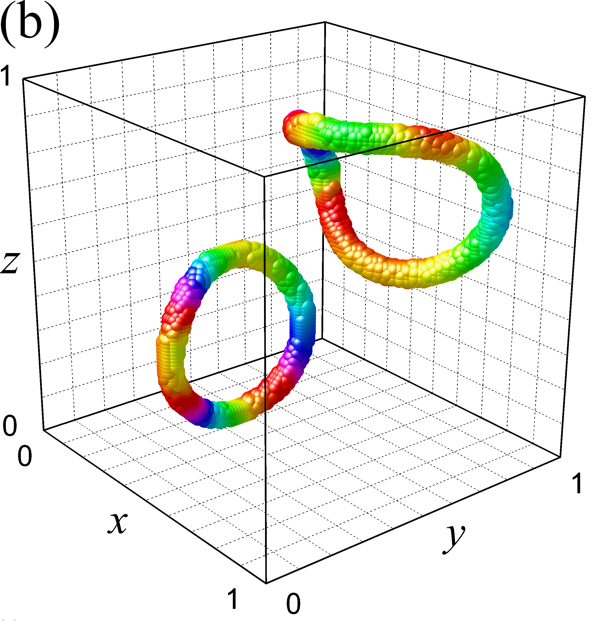} \   \ 
  \includegraphics[width=0.22\linewidth]{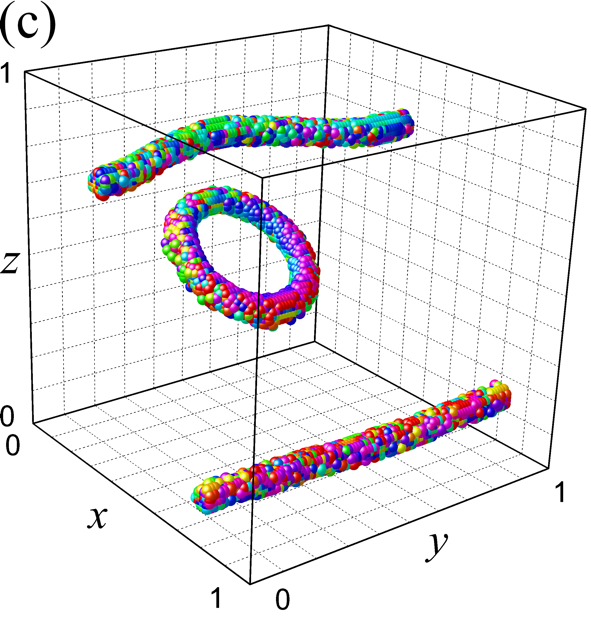} \ \   
  \includegraphics[width=0.22\linewidth]{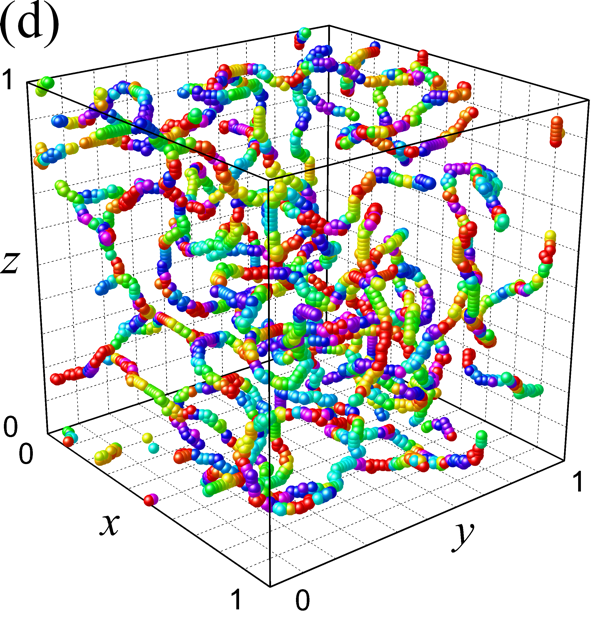}  \ \
  \includegraphics[width=0.035\linewidth]{Faza-insert.png}  \\
  \includegraphics[width=0.215\linewidth]{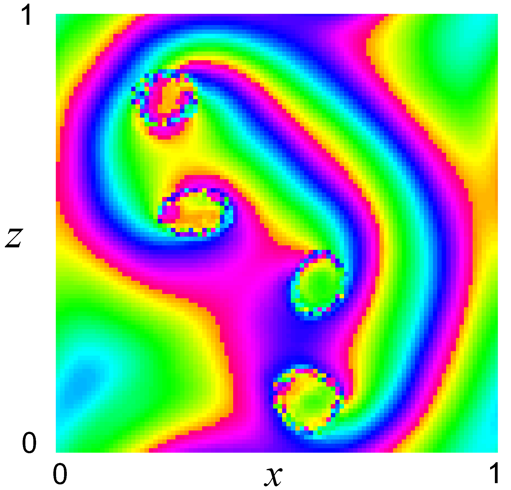}  \   \  \ 
  \includegraphics[width=0.215\linewidth]{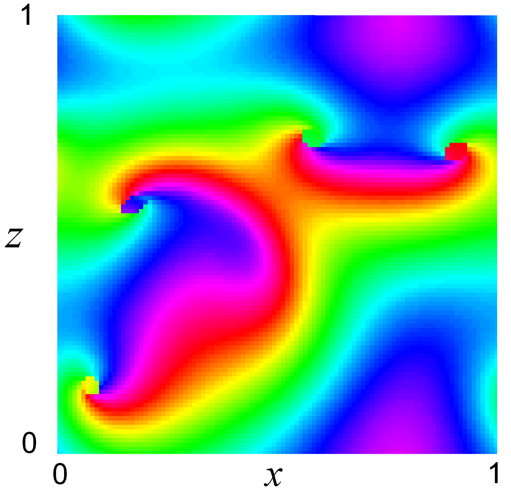} \   \  \  \ 
  \includegraphics[width=0.215\linewidth]{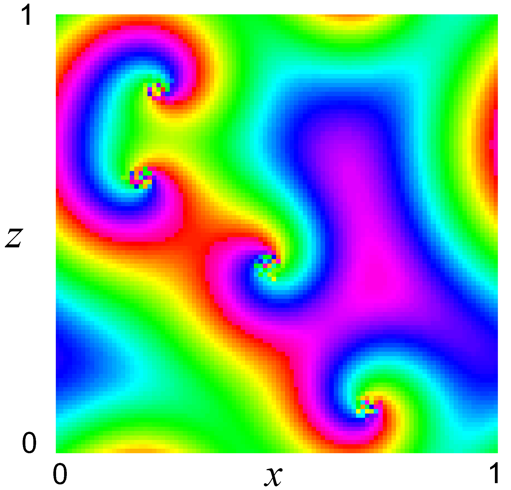}  \  \  \  \  
 \includegraphics[width=0.215\linewidth]{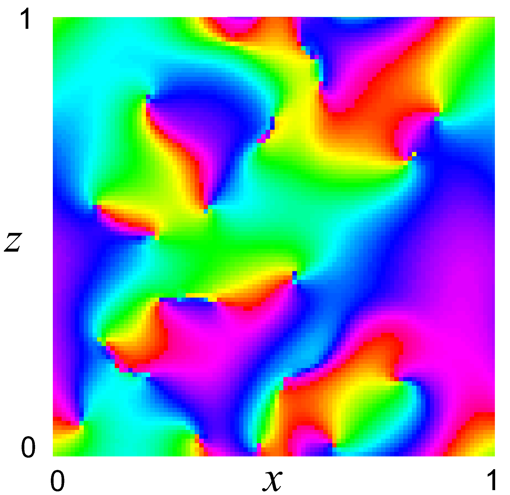}  \ \   
  \includegraphics[width=0.035\linewidth]{Faza-insert.png}} \
\vspace*{-0.4cm}
 \caption{ Scroll wave chimeras in system (1). Phase snapshots and their cross-sections: (a) -  trefoil ($ \alpha=0.35, \epsilon=0.05, \mu=0.1, r=0.03, N=100$);  (b) - two rings  ($\alpha=0.44, r=0.04,\mu=0.048, \epsilon=0.052, N=100$). (c) - ring with two rods  ($\alpha=0.59, r=0.055,\mu=0.01, \epsilon=0.1, N=100$);  (d) - multiple scroll wave chimera  ($\alpha=0.2, r=0.02,\mu=0.1, \epsilon=0.05, N=100$). 
 Coordinates  $x=i/N,  y=j/N,  z=k/N$. } 
\label{fig:5}
\end{figure*}

Our simulations demonstrate that though  Fig. 3 is obtained for model (1) with $N=100$, but similar parameter regions exist for larger $N = 200$ and the same relative coupling radius $r = P/N$. 

The examples of scroll ring chimeras with incoherent, partially coherent,
and completely coherent inner parts are presented in Fig. 4. Location of parameter values for the examples are indicated by black points in Fig. 3(a).
The average frequency profile of an incoherent scroll ring  in  Fig. 4(a)  \href{video 2.avi} {(Multimedia view)} is smooth and bell-shaped, as typically
occurs for chimeras in the Kuramoto model without inertia. Figure 4(b) illustrates  the scroll ring chimera with partially coherent inner part and incoherent  boundary. 
In the case of a completely coherent ring (Fig. 4 (c)), we have a bistable behavior of oscillators. 
Their average frequencies $\bar{\omega}_{xyz}$ are equal to the average frequency of the ring $\bar{\omega}_{CR}$ or  
main synchronized cluster $\bar{\omega}_{0}$ only. So, it is a scroll ring pattern, but is not  scroll ring chimera. 

Figure 4 illustrates three average frequency profiles of the inner parts of the scroll rings only. But we suggest that, by increasing the dimension $N$ of system (1), the different scroll ring  structures including multiple stepwise structures similarly to the 2D case ~\cite{msm2019} can be obtained.

To our surprise,  if we take a scroll ring with the coupling radius $r=0.041$ and $\alpha=0.3$ (yellow point in Fig. 3(b)) and will start to increase the coupling radius $r$ along the black dashed line, the scroll ring  is transformed into a sphere pattern without spiral rotation (shown in the insert in Fig. 3(b)). 
Approaching the left and bottom boundaries of the stability region of the sphere (Fig. 3(b), the number of incoherent oscillators on the surface decreases. Finally, the sphere vanishes, by crossing the boundaries of this region.
 Near the right boundary of the stability region, the sphere becomes an incoherent ball.  
Just after crossing the right boundary, the ball is transformed firstly into a cube and finally into the pattern consisting of incoherent oscillators with coherent islands which fill all the space except for a narrow coherent layer (shown in the insert in Fig. 3(b)).

\begin{figure}[ht!]
  \vspace*{-0.6cm}
 \center{\includegraphics[width=0.42\linewidth]{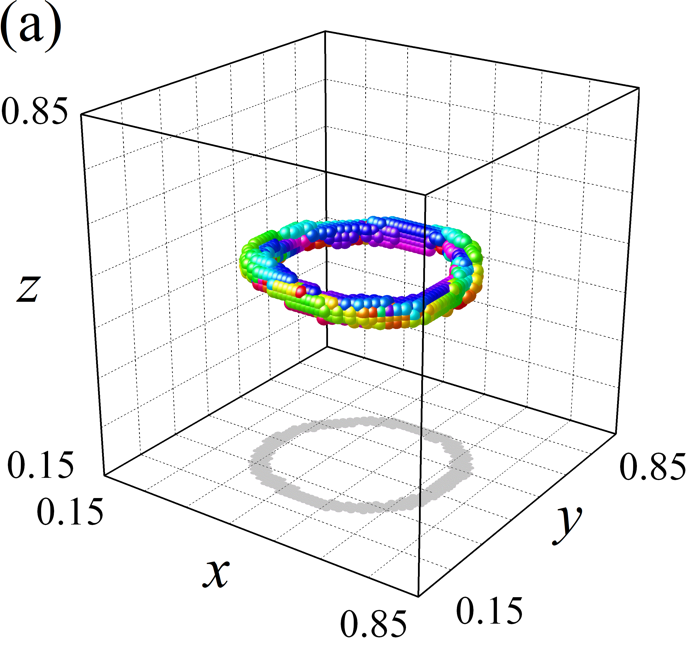}  \  
  \includegraphics[width=0.42\linewidth]{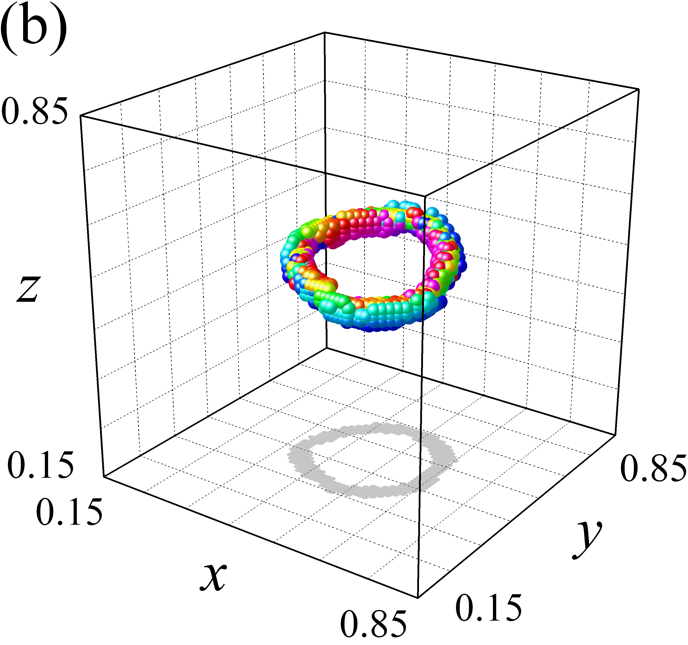}   \
  \includegraphics[width=0.06\linewidth]{Faza-insert-Fig2.png} \\
  \includegraphics[width=0.42\linewidth]{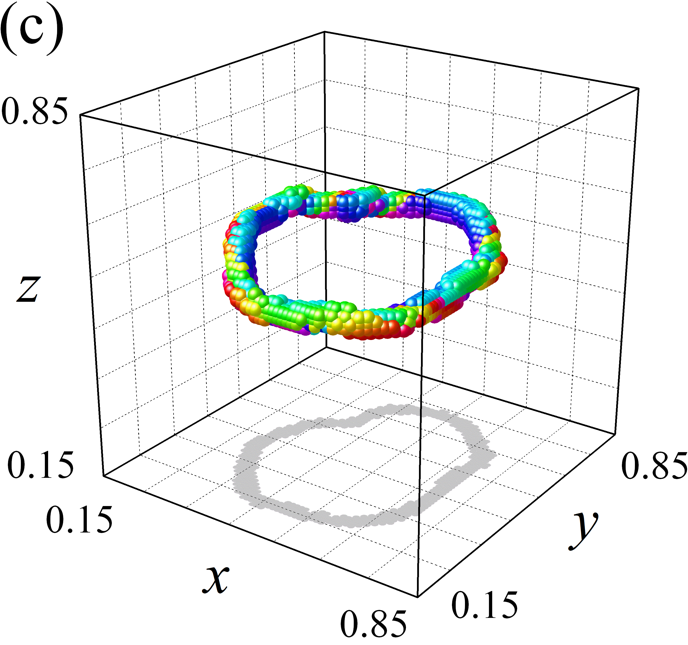} \   
  \includegraphics[width=0.42\linewidth]{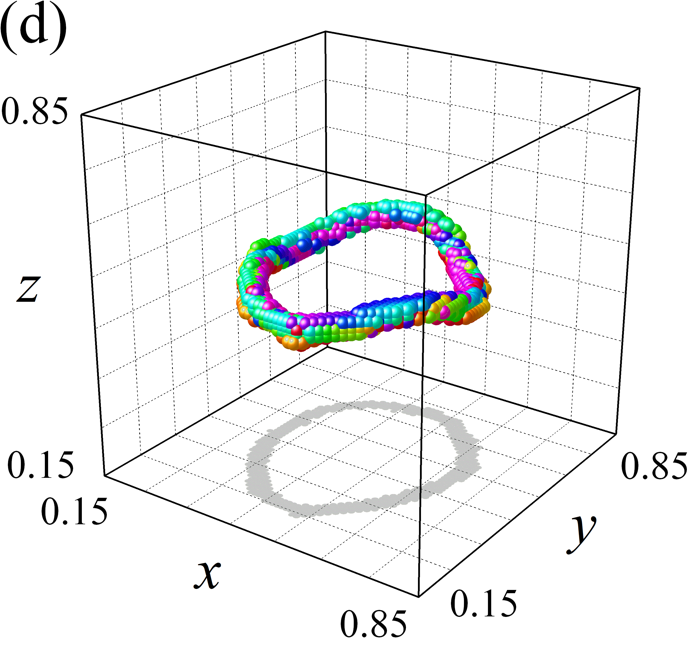}  \  
  \includegraphics[width=0.06\linewidth]{Faza-insert-Fig2.png}} 
  \vspace*{-0.3cm}
\caption{Perturbated of the scroll ring chimera.  (a) - origin scroll ring chimera before perturbation,  (b-d) - perturbated scroll ring chimera  with amplitude $0.8$.  $\alpha=0.38, r=0.04, \epsilon=0.05, \mu=0.02, N=100$. Simulation time $t=10^4$. } 
\label{fig:6}
\end{figure}

\begin{figure*}[ht!]
  \vspace*{-0.6cm}
 \center{\includegraphics[width=0.22\linewidth]{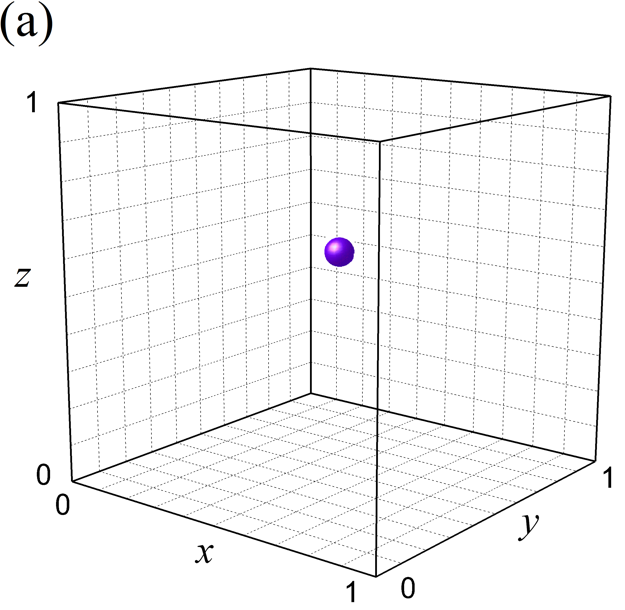}  \hspace*{0.6cm} 
 \includegraphics[width=0.03\linewidth]{Faza-insert.png} \hspace*{0.4cm} 
   \includegraphics[width=0.215\linewidth]{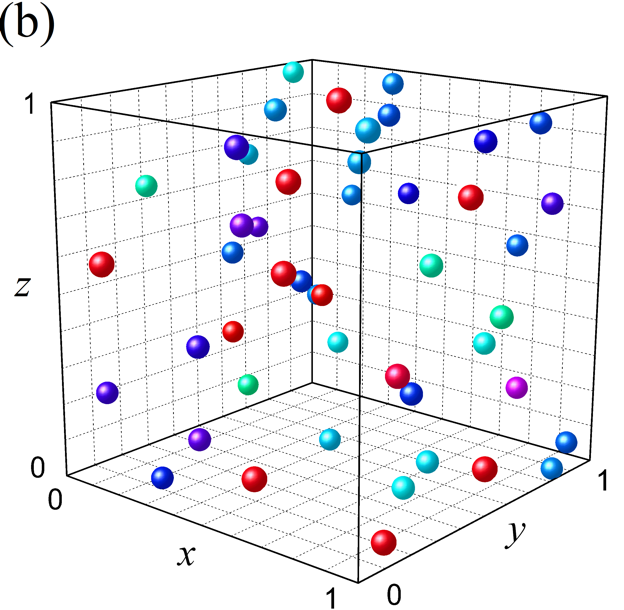}   \hspace*{0.6cm} 
  \includegraphics[width=0.36\linewidth]{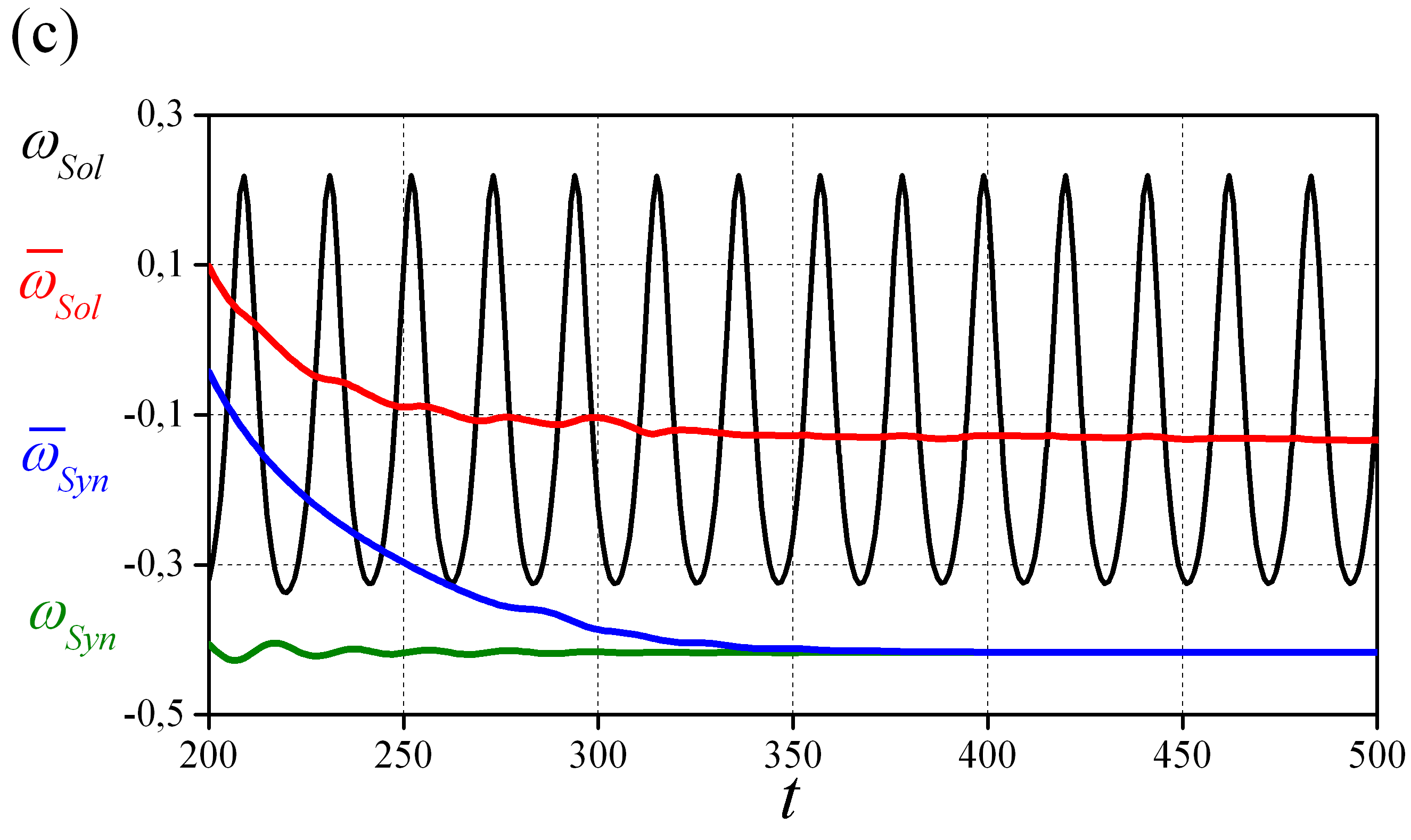} } 
\vspace*{-0.4cm} 
 \caption{ Solitary states. (a) -  phase snapshot for a single solitary state ($\alpha=0.21$), (b) -   phase snapshot  for multiple solitary states ($\alpha=0.25$), (c) - time evolution of frequencies for a single solitary state. 
$ \mu=0.1, \epsilon=0.05, r=0.04, N=50$. Frequency averaging interval $\Delta T = 200$. 
 Coordinates  $x=i/N,  y=j/N,  z=k/N$. } 
\label{fig:7}
\end{figure*}

\begin{figure*}[ht!]
\vspace*{-0.3cm} 
\includegraphics[width=0.218\linewidth]{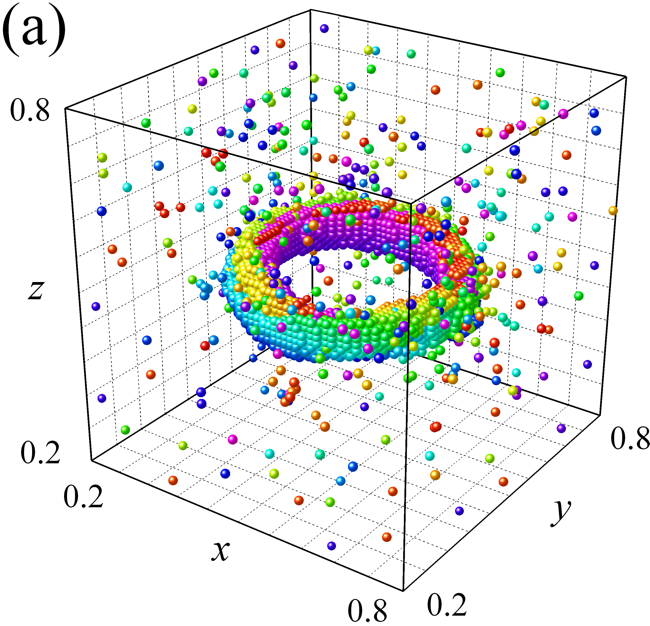}     \hspace*{0.19cm} 
\includegraphics[width=0.21\linewidth]{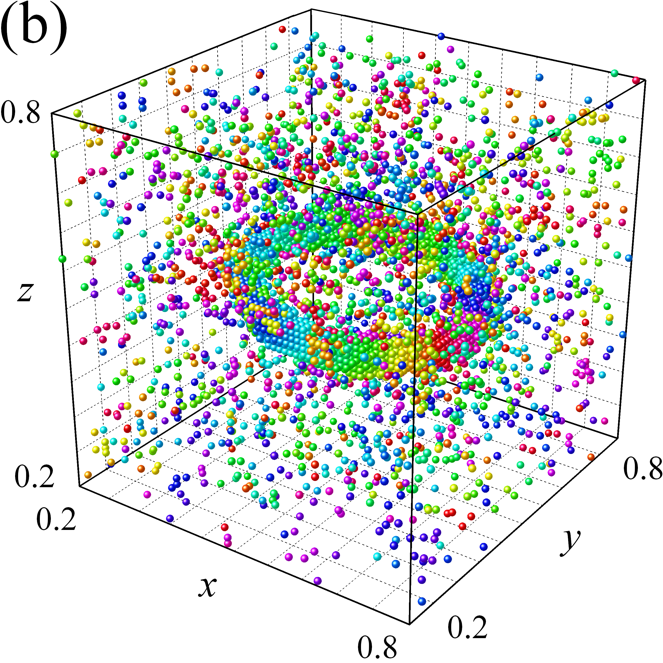}   \hspace*{0.2cm}
\includegraphics[width=0.218\linewidth]{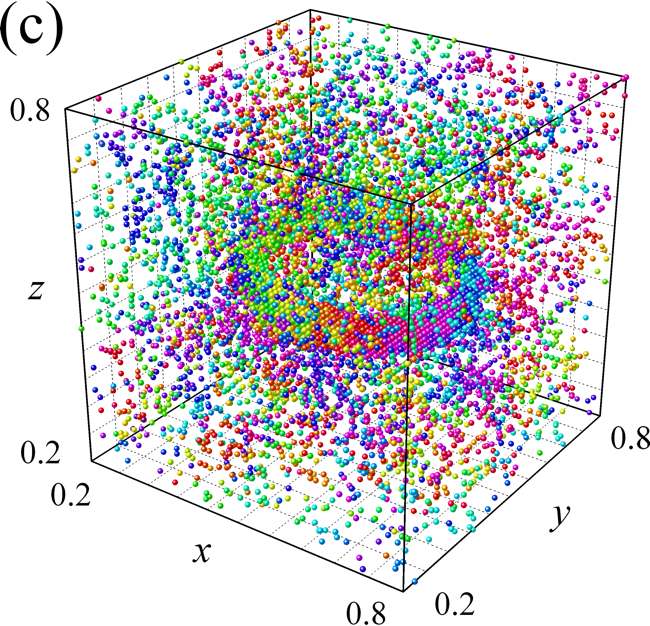}    \hspace*{0.2cm}
  \includegraphics[width=0.218\linewidth]{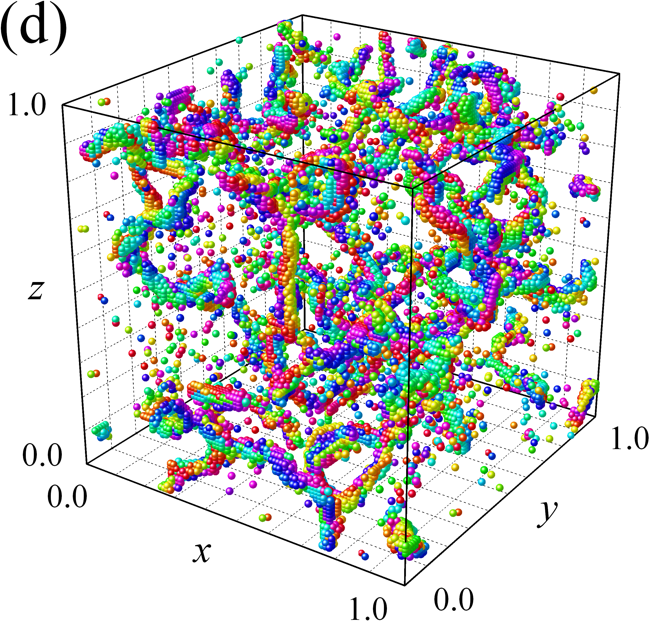}   \hspace*{0.2cm}
  \includegraphics[width=0.03\linewidth]{Faza-insert.png}  \\
\vspace*{0.2cm}
\includegraphics[width=0.212\linewidth]{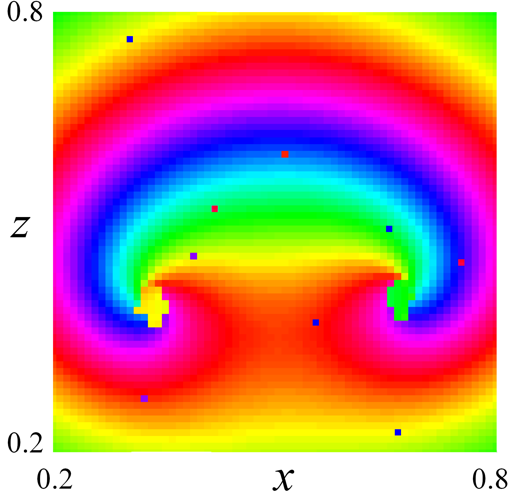}   \hspace*{0.28cm}
\includegraphics[width=0.212\linewidth]{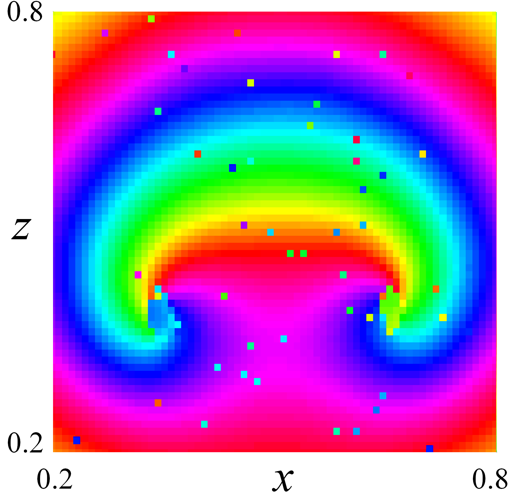}   \hspace*{0.3cm} 
\includegraphics[width=0.212\linewidth]{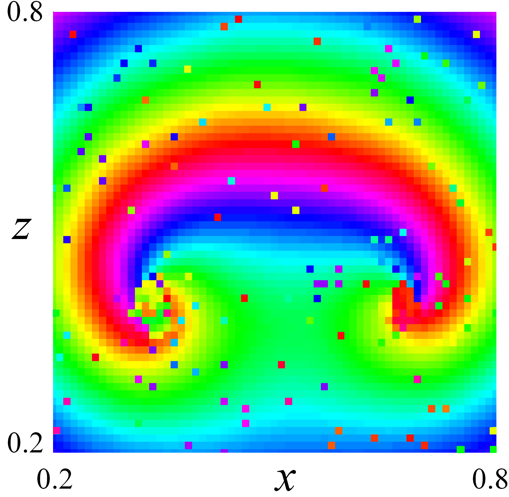}   \hspace*{0.31cm}
  \includegraphics[width=0.212\linewidth]{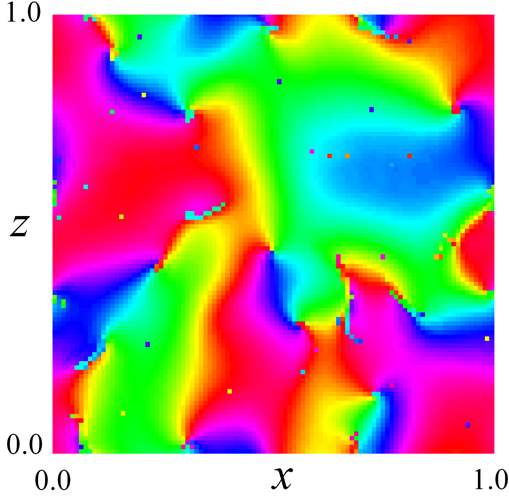}  \hspace*{0.24cm}
  \includegraphics[width=0.032\linewidth]{Faza-insert.png} 
 \vspace*{-0.3cm}
\caption{Examples of the scroll wave chimera states with solitary clouds.  Phase snapshots  and their cross-sections. 
Scroll rings chimera perturbed  with amplitude:
  $0.5$ (a),   $0.55$   (b),   $0.6$ (c) ($\alpha=0.4, \mu=0.1, \epsilon=0.05, r=0.03, N=100$).  (d) - multiple scroll wave chimera obtained from random initial conditions ($\alpha=0.22, \mu=0.1, \epsilon=0.05, r=0.028, N=100$).}
  \label{fig:8} 
\end{figure*}

 In the system without inertia, the scroll wave chimeras have, obviously, incoherent inner parts \cite{msom2015,ld2016,msom2017}.  But, in the system with inertia, as one can see from the phase cross-sections of chimeras in Fig. 5,  they  may have coherent or partially coherent inner parts. 
Due to the introduction of the inertia, the
trefoil  (Fig. 5(a) and Hopf link chimeras can be fixed in the oscillatory space in contrast to their behavior in a system without inertia, where they move, by rotating around their center of mass. 

Scroll wave chimera states from system (1) are very stable with respect to perturbations of the initial conditions. 
For example, the scroll ring chimera still exists and retains its shape, even if its phase $\varphi_{xyz}$ and frequency ${\omega}_{xyz}$ are perturbed by uniformly distributed noise with amplitude less than $0.5$ at the parameter values $\alpha=0.38, r=0.04, \epsilon=0.05, \mu=0.02, N=100$ (yellow point in Fig. 3(a)). Stronger perturbations lead to changing the shape of chimeras or its destruction  with complete oscillatory synchronization or creation of different types of scroll wave chimera states such as: a few rings, rods, etc.
If we take a scroll ring (Fig. 6(a)) outside the solitary region
and start to perturb it by uniformly distributed noise  with amplitude $0.8$, then we can
 obtain  various new scroll ring chimera states with different shapes (Fig. 6(b-d)). So, in this way, we can generate many different rings for fixed parameter values of system (1). 
For each chimera obtained in this way, we can again apply the perturbation method to ganerate new  scroll ring chimera states and so on. 

The dynamics of system (1) becomes more complicated, if the parameter values enter  into  the solitary region ( hatched in gray
 in Fig. 3) where isolated  oscillators  exist~\cite{JMK2015,JBLDKM2018}. 
Single (Fig. 7(a)) or multiple (Fig. 7(b)) solitary states can be easily obtained from random initial conditions. Multiple solitary states look like a "solitary cloud''.
An example of the time evolution of frequencies ${\omega}_{xyz}$ for a single solitary state from Fig. 7(a)  is shown in Fig. 7(c).  
The frequency of a solitary state ${\omega}_{Sol}$ becomes very soon periodic, and its average value 
$\bar{\omega}_{Sol}$ tends to a constant. 
The frequencies of synchronized oscillators ${\omega}_{Syn}$ and their average values 
$\bar{\omega}_{Syn}$ tend to constants as well. 
If the disturbed chimera lies in the solitary region, then the perturbation of its initial conditions  with amplitude in the interval  $(0.5 - 0.8)$ can
give rise to another chimera states with solitary clouds. 
Examples of  scroll ring chimeras   with solitary clouds  are shown in Fig. 8 for the parameter values indicated by the green point in Fig. 3(a).
As clearly seen from these figures, the solitary clouds in the oscillatory space can appear with a perturbation  of a scroll ring with amplitude $0.5$ (Fig. 8(a)). Stronger perturbations with amplitudes $0.55$  (Fig. 8(b))  and  $0.6$  (Fig. 8(c)) save the scroll ring chimera, but clouds becomes denser. The further perturbation of chimera's initial conditions  with amplitude more than $0.8$ can destroy the original chimera states.
We show  also multiple scroll wave chimera with solitary clouds (Fig. 8(d)) obtained from random initial conditions. 

 In such a way, any scroll wave chimera from the solitary region with fixed parameters  may have  solitary clouds of arbitrary  random shape. Such clouds may be obtained by a random perturbation of the original chimera. 
However, the transition  between solitary and non-solitary domains  is noninvertible. 
 Perturbations  of  chimeras with parameters outside the solitary regions  never  lead  to the formation of solitary clouds.  At the same time,  starting from initial conditions with solitary clouds  outside the solitary region leads to the vanishing of solitary clouds.  On the other hand, starting from initial conditions without solitary cloud inside the solitary region  never  leads to the formation of a solitary cloud.

 \begin{wrapfigure}[12]{r}{0.46\linewidth} 
\vspace{-1.0cm}
\begin{center}
\includegraphics[width=0.8\linewidth]{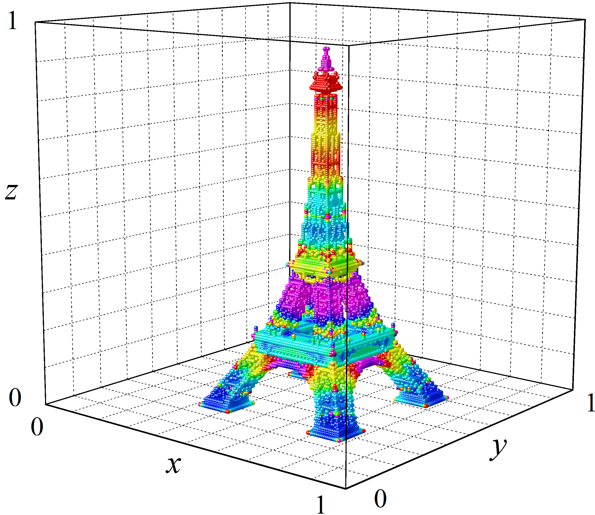}  \    
  \includegraphics[width=0.09\linewidth]{Faza-insert.png}
\vspace*{-0.4cm} 
 \caption{Phase snapshot of 3D image of the Eiffel Tower  as a solution of the  3D system with inertia (1) \href{video 3.avi} {(Multimedia view)}.. Parameters $ \alpha=0.4, r=0.2, \mu=0.1,  \epsilon=0.05, N=200$.}
  \label{fig:9} 
\end{center}
 \end{wrapfigure}

Finally, we would like to propose a method of construction of 3D images using solitary states as the solutions of the  3D Kuramoto model with inertia (1). 
In addition to the image of a chimera sculpture from the Notre-Dame Cathedral (Fig. 1(d)), we present  the Eiffel Tower image in  Fig. 9 \href{video 3.avi} {(Multimedia view)}.

This image is a stable solution of the 3D Kuramoto model with inertia (1). 
Stable solitary patterns of arbitrary shape may be constructed  using  dynamic variables of synchronized and solitary oscillators from any other stable solitary state.
A model for 3D printing  was used  as a template  for the spatial placing of  solitary oscillators.

The 3D Kuramoto model with inertia (1) describes the basic properties of the collective dynamics in various real physical systems.  Equations similar to (1) arise in magnonics \cite{t2014,klti2017}, optics \cite{fcr2018}, etc. The phenomena similar to chimera and solitary states were observed experimentally in various systems, e.g., with the Josephson effects,  Bose--Einstein condensation of magnons \cite{t2014,klti2017}, pulling of frequencies and optical modes \cite{fcr2018}, etc.  
Models of oscillating nonlinear networks are used in biology and electronics. Such properties as the stability of the described chimera and solitary states with respect to perturbations of the initial conditions in a wide range of parameters and the preservation of initial average oscillating frequencies for solitary states may become a key to practical applications. 

The practical applications of these phenomena can include the creation of new information storage and transfer media, devices, architectures; development of new information encoding algorithms; clarification of the mechanisms of functioning of biological systems, etc. 

The authors are very grateful to  Yu. Maistrenko  for useful  discussions and valuable comments.

{\bf Data availability.}
The data that support the findings of this study are available from the corresponding author (maistren@nas.gov.ua) upon reasonable request.

\end{document}